\documentclass{elsart}
\usepackage{graphicx}
\usepackage{amssymb}
\usepackage{url}
\begin{document}
\begin{frontmatter}

\title{Parallel Distribution of Asynchronous Optical Signals}

\author[leeds]{RJ White\corauthref{cor}},
\ead{rw@ast.leeds.ac.uk}
\author[leeds]{HJ Rose},
\author[leeds]{SM Bradbury},
\author[hytec]{P Marshall}
\address[leeds]{School of Physics and Astronomy, University of Leeds, Leeds, LS2 9JT,UK}
\address[hytec]{Hytec Electronics Ltd., Reading, UK}
\corauth[cor]{Corresponding author.}


\begin{abstract}
An eleven channel digital asynchronous transceiver (DAT) employing
parallel optical link technology has been developed for trigger signal
distribution across the Very Energetic Radiation Imaging Telescope
Array System (VERITAS). Combinatorial logic functions are implemented
in Xilinx Spartan 3 FPGAs, providing a versatile solution adaptable
for use in future atmospheric \v{C}erenkov detectors and other
high-energy astroparticle experiments. The device is dead-time free
and introduces a minimal skew of \mbox{1.6 ns} between channels. The
jitter on each DAT channel is less than \mbox{0.8 ns} 95\% of the
time, allowing communication between telescopes and a central trigger
system separated by hundreds of meters, without limiting array
performance.
\end{abstract}
\end{frontmatter}

\section{Introduction}
\label{Introduction}
The Very Energetic Radiation Imaging Telescope Array System (VERITAS)
\cite{Holder06} consists of four \mbox{12 m} diameter atmospheric
\v{C}erenkov telescopes equipped with photo-multiplier cameras. The
array uses a multi-level trigger system to reject fluctuations in
background light whilst efficiently recording signals from gamma-ray
initiated air showers. Discriminated photo-multiplier signals are
passed to a pattern selection trigger at each telescope
\cite{Bradbury02} resulting in a cosmic ray trigger rate at high
telescope pointing elevations of $\sim$150 Hz for a discriminator
threshold of $\sim$6-7 photoelectrons. These decisions are then passed
to a central array trigger requiring a multiple telescope
coincidence. The array trigger requires the distribution of multiple,
fast, digital pulses between telescopes. A versatile solution is
essential as not only fast, narrow, trigger signals but also variable
width event numbers and long calibration flags must be transmitted.

\begin{figure*}[t]
\begin{center}
\includegraphics*[angle=90, width=1.0\textwidth]{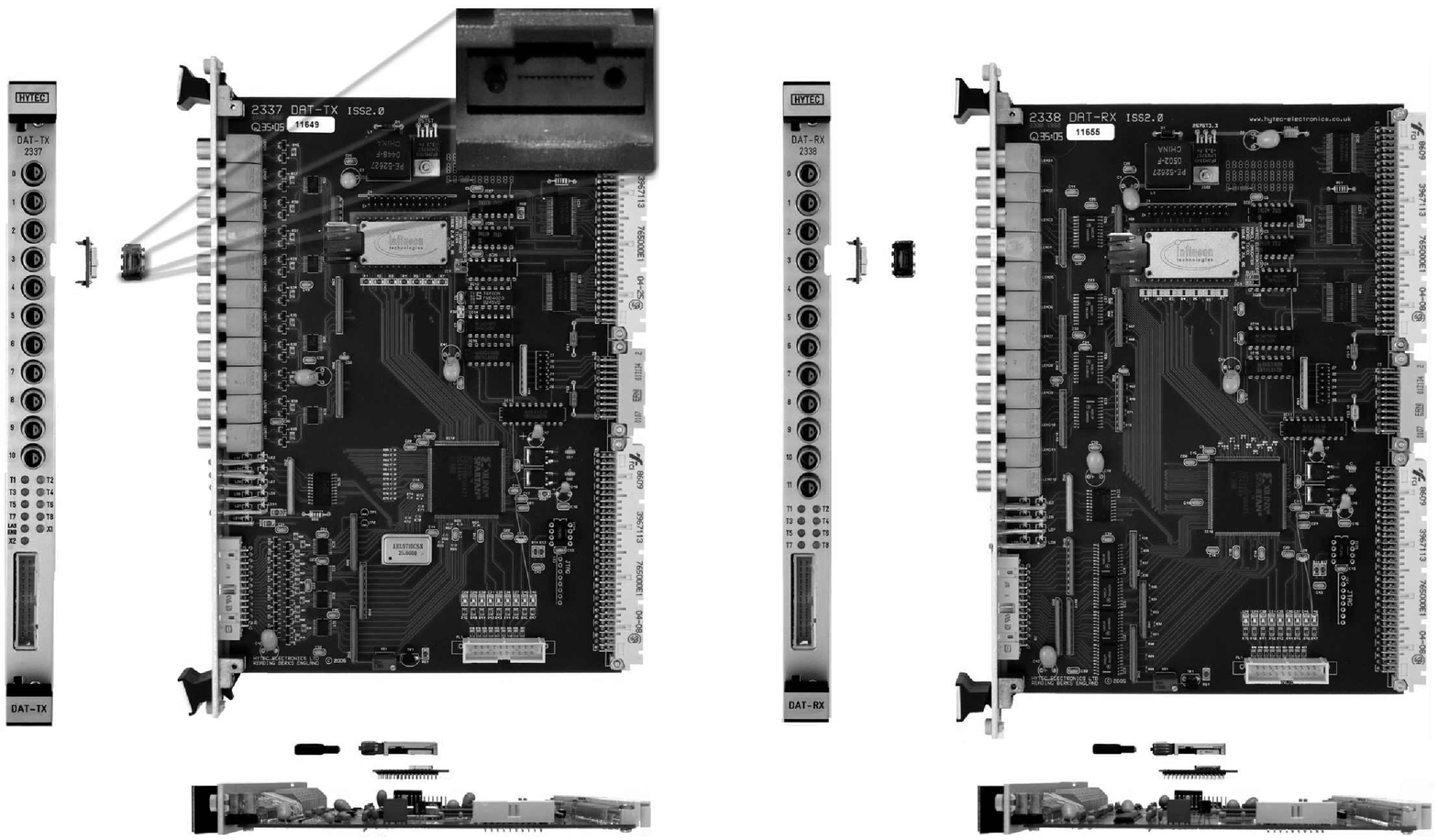}
\end{center}
\caption{The first DAT pair, transmitter on the left, receiver on the
right, with a close up of the Infineon PAROLI \cite{paroli} inset.}
\label{fig:boards}
\end{figure*}

The data from each camera pixel is digitised into 2 ns slices by
custom built \mbox{500 MHz} FADC boards \cite{Buckley03}. Upon
receiving an array trigger signal the FADC buffers are readout. The
trigger signals require nanosecond accuracy to trigger the readout of
the FADC modules from the correct point in the buffers. The FADC
readout time has a direct effect on the telescope deadtime and energy
threshold, so it is important the minimise the readout window around
the data pulse. A large amount of jitter in the trigger signal would
require a large readout window, increasing the system deadtime and the
detector energy threshold. Typically the FADC readout window is set to
24 samples, or \mbox{48 ns} in VERITAS. Accordingly a Digital
Asynchronous Transceiver (DAT) has been developed at the University of
Leeds in collaboration with Hytec Electronics Ltd. consisting of a
transmitter and a receiver both implemented as single width VME
modules, Figure~\ref{fig:boards}. The transmitter and receiver are
linked by fibre optic interconnects running between telescopes. Fibre
is lighter and less bulky than coaxial cable of the equivalent length
and is not susceptible to electromagnetic interference.

\section{Specification}
\label{Specification}
The DAT transmitter must accept differential negative emitter coupled
logic (NECL) inputs via both an IDC header and twin-axial LEMO
connectors before presenting them in an identical form at the receiver
outputs. Ideally, the solution should exceed the current trigger
requirements and be adaptable for future developments, both within
atmospheric \v{C}erenkov astronomy and high energy astrophysics in
general.

To minimise the FADC readout window within VERITAS it is desirable to
maintain the trigger timing to the level of one FADC sample. The array
trigger signal distributed by the DAT modules is based upon telescope
level triggers also transmitted over DAT modules and received
centrally. The array trigger itself has an RMS jitter of \mbox{$\sim$1
ns}, so to keep the total RMS jitter of the array trigger signal below
one FADC sample, each channel of a DAT module should have \mbox{$<$1
ns} RMS jitter. The array trigger system uses a pulse delay module to
compensate for the differences in timing introduced by the difference
in arrival time of the shower front at the telescopes depending on the
observation direction. This method requires the array system signals
traverse a path of predictable duration. Therefore a maximum
channel-to-channel skew of \mbox{2 ns} within the DAT modules is
desired.

For versatility, the DAT must transport asynchronous pulses of varying
width at an undefined frequency such that the same system can be used
to distribute housekeeping information\footnote{In the case of VERITAS
the width is of the order of tens of nanoseconds to milliseconds
at frequencies of the order of Hz up to a few MHz.}. The system should
be deadtime free to remove the possibility of missing incoming pulses
whilst the data is being transmitted.

\section{Implementation}
\label{Implementation}
\subsection{Overview}
\label{Implementation_Overview}

\begin{figure}[t]
\begin{center}
\includegraphics*[angle=90, width=1.0\textwidth]{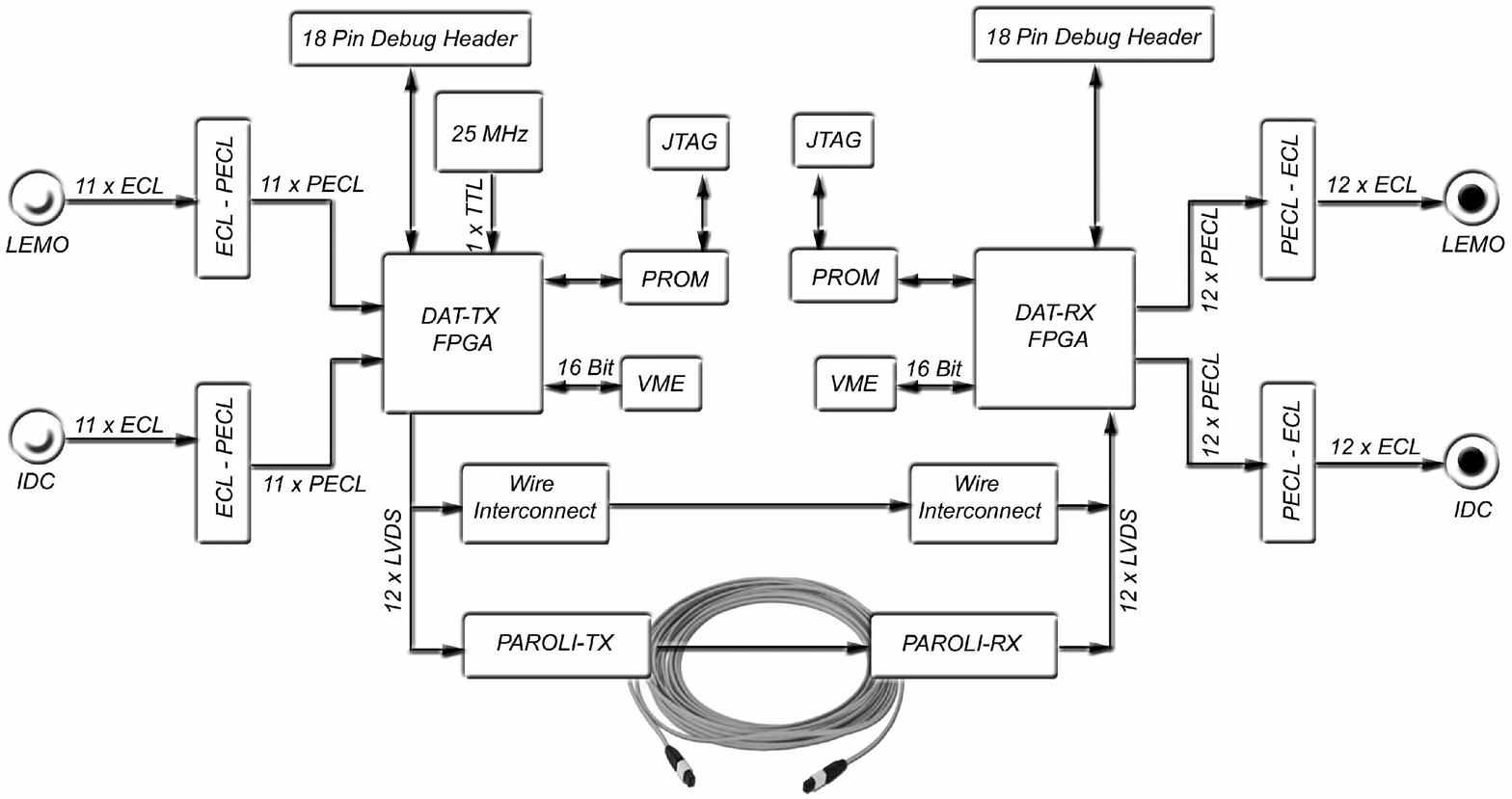}
\end{center}
\caption{Functional diagram of the DAT modules depicting the data flow
from transmitter to receiver including signal standards. Either fibre
or wire interconnect may be selected. The 12 channel MPO terminated
fibre cable is shown inset.}
\label{fig:functionaldiagram}
\end{figure}

The DAT consists of two, 6U high, single width VME modules denoted
\mbox{DAT-TX} and \mbox{DAT-RX}, transmitter and receiver
respectively, see Figure~\ref{fig:functionaldiagram}. As described in
Section~\ref{Implementation_FPGACode}, control and monitoring takes
place via a VME interface embedded into an onboard Xilinx
\mbox{Spartan 3} XC3S50 Field Programmable Gate Array (FPGA)
\cite{spartan3}.

To maintain signal integrity and avoid any lightning induced power
surges, which are a hazard with copper cable, a \mbox{62.5/125} $\mu$m
core fibre optic interconnect is used to transport signals over the
distance of at least \mbox{150 m} between telescopes. The conversion
of electrical signals to optical signals is achieved using the
Infineon \mbox{1.25 Gbit/s} parallel optical link (PAROLI
2$^{\tiny{\textregistered}}$ \cite{paroli}, see
Figure~\ref{fig:boards}) consisting of a small form factor, 12
channel, \mbox{850 nm} VCSEL driven transmitter and PIN diode array
based receiver. The PAROLI laser devices have a range of around 800 m
that puts an upper limit on the distance between DAT modules and
therefore telescopes. As a matter of laser safety a given PAROLI
transmitter channel will become disabled if a signal exceeds a duty
cycle (DC\%) of 57\% within \mbox{1 $\mu$s}. Thus the asynchronous
variable width input data signals {\it cannot} be sent to the PAROLI
directly, they must first be modulated to the DC\% requirement in a
recoverable manner.

The minimum signal switching time and DC\% condition of the PAROLI are
satisfied by encoding incoming data signals with a 25 MHz clock
through exclusive OR (XOR) gates at the transmitter. The encoded data
signals and a copy of the clock are optically transmitted to the
receiver where a second set of XOR gates recovers the data. Using this
asynchronous, combinatorial method deadtime incursions associated
with sequential logic are avoided.

Combinatorial logic is usually implemented in surface mounted,
function-specific, IC chips. To accommodate rapid design change
(e.g. to optimise the boards for additional applications) and the
future possibility of a PAROLI without rigorous duty cycle
constraints, all logic is performed within the onboard FPGA. As the
gate array is used to implement the VME interface this is an
economical solution.

\subsection{Physical Description}
\label{Implementation_PhysicalDescription}

Differential NECL data enters the front of the DAT-TX via either the
26 way IDC header or the eleven twin-axial LEMO connectors, selectable
over the VME interface.

Surface mounted, three channel, MC10EP90 \cite{MC10EP90} chips convert
NECL into low-voltage positive emitter coupled logic (LVPECL) suitable
for input to the FPGA. Retaining a differential signal standard
improves signal integrity but requires twice as many FPGA input
pins. The 25 MHz clock is provided by an onboard TTL oscillator.

The XOR encoded data and clock are output from the FPGA in low voltage
differential signal (LVDS) form; the required input to the PAROLI. The
PAROLI itself is attached to a daughter board using a 100 pin, ball
gate array mounted connector. This configuration would allow the
PAROLI to be replaced with a future device by simply redesigning the
daughter board. A male MPO terminated 12 channel ribbon-fibre is
plugged directly into the PAROLI.

At the receiver the PAROLI converts the incoming optical signals
back to LVDS. There is also a 26-way wire interconnect option
available to communicate between DAT modules during system
debugging. The LVDS signals are then input to the receiver's Xilinx
Spartan 3 where the data is recovered.

Recovered data lines are output from the FPGA as LVPECL and converted
with three channel MC100EL91 \cite{MC100EL91} chips to NECL. The eleven
data signals and clock signal are presented on the front panel to both
twin-axial LEMO connectors and a 26 way IDC header.

Upon power-up the FPGA on each module is programmed from a socket
mounted Atmel AT17LV512 EePROM \cite{prom} via jumper selection. In
the absence of a PROM programming can be done via a JTAG header. This
setup provides the most versatile programming architecture, allowing
the user to reprogram the FPGA in seconds over the JTAG connection
when testing modifications to the modules.

\subsection{FPGA Combinatorial Encoding and Decoding}
\label{Implementation_FPGACode}

The embedded VME interface register set provides 16 bit access to
three registers. The first two registers simply read the module model
number and ID. The third acts as a control status register (CSR) and
allows the user to control and monitor the modules. The CSR can enable
and disable the laser used in the PAROLI and the clock used in the XOR
encoding. The transmitter input choice of IDC header or twin-axial
LEMO connector is also set using the CSR, and a user LED on the front
panel may be controlled for diagnostic purposes. A test header on the
boards may be enabled to input and output 7 additional signals to the
FPGA. These pins may be assigned a function and signal standard by
reprogramming the FPGA. A simple C++ program scans the VME crate and
records the slot numbers of all DAT modules present, identifying them
as either transmitter or receiver. The user may then select a given
module to access the CSR. Automated programs scan and start up or shut
down all the modules in a crate, enabling or disabling the laser and
clock, and setting the LVDS inputs via a configuration file or, in the
case of VERITAS, a database table. At present the status of
transmitted data may not be monitored over the VME interface, and
equally data pulses may not be injected in this way. Whilst this
functionality is straightforward to implement tests showed that
changing the FPGA code to intercept the data paths effected the
timing of the signals which, as described later in this section, is
crucial to the correct recovery of the data. Further VME functionality
is desirable in any future iterations.

At the transmitter, the 22 differential LVPECL signals from both the
twin-axial and IDC inputs are buffered onto the FPGA. Each input is
selected individually over VME via multiplexers within the gate array.
The 25 MHz clock is input via a dedicated clock-buffer to a low skew
network on the FPGA and fanned out 12 times. Each of the fanned out
clocks enters an XOR gate with a corresponding data line, except for
the twelfth clock which enters an XOR gate with ground to maintain
duty cycle and transit time. The eleven encoded data lines and clock
encoded with ground are buffered to differential outputs on the
FPGA. During laboratory tests the duty cycle of the encoded
transmitter output was seen to vary by around \mbox{0.5\% (200 ps)}
from a low input data state to a high input data state. This duty
cycle dependence on the high/low state of the input is attributed to
rising and falling signal edges taking different paths through the
FPGA, leading to unequal transition times.

\begin{figure}[t]
\begin{center}
\includegraphics*[angle=-90, width=1.0\textwidth]{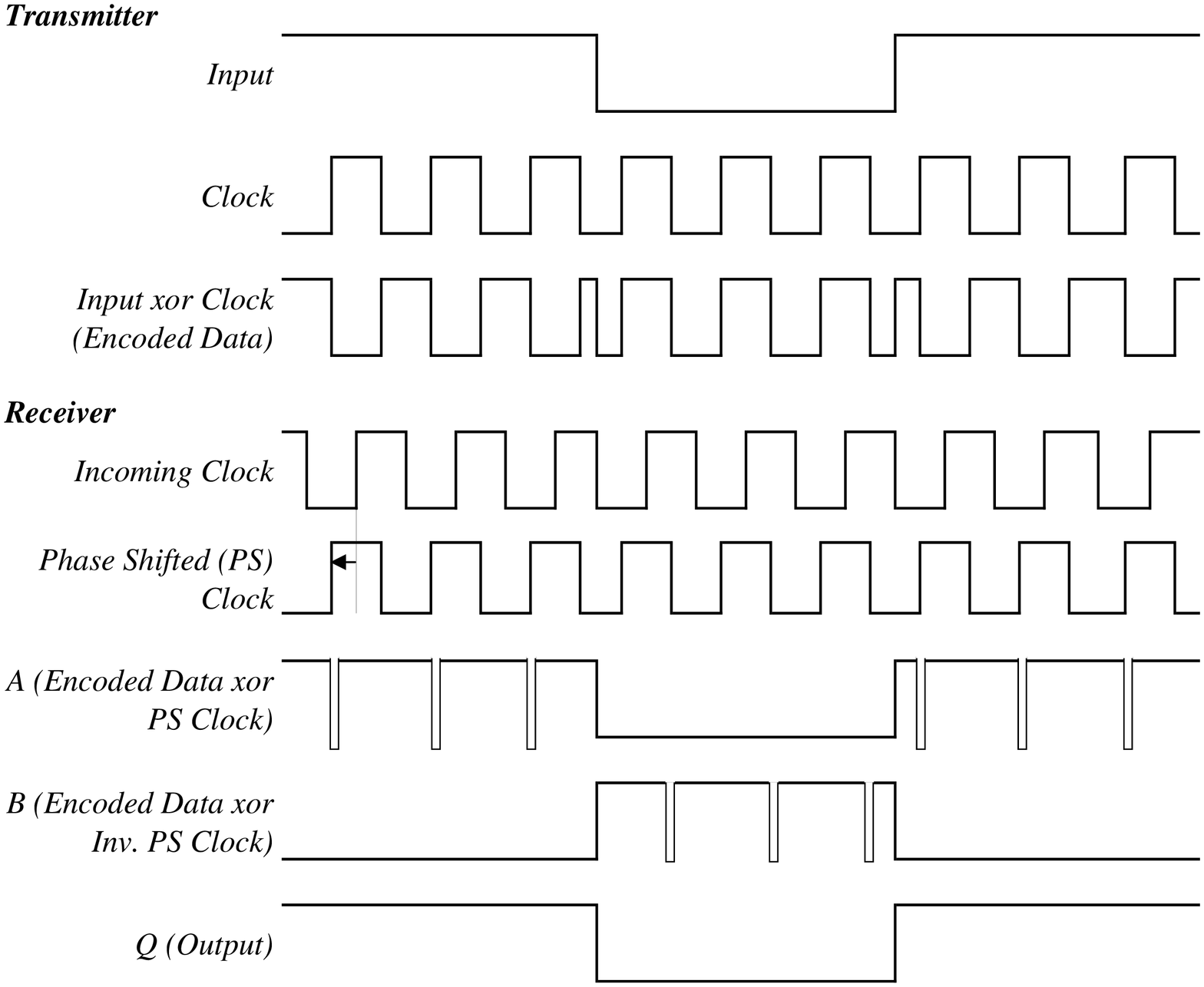}
\end{center}
\caption{Depiction of XOR encoding and decoding, embedded in the DAT
FPGA chips.}
\label{fig:xor}
\end{figure}

At the receiver the differential clock is buffered to a digital clock
manager (DCM). Feedback into the DCM facilitates a phase shift between
the input and output with a resolution of \mbox{156 ps} for a \mbox{25
MHz} clock. The phase shifted clock is fanned out and entered along
with the 11 data lines to XOR gates. As previously noted, the duty
cycle of the encoded signal exiting the transmitter and arriving at the
receiver depends on whether the input to the transmitter is high or
low. Due to this duty cycle dependence on the high/low state of the
input it is not possible to phase shift the clock to a point where the
input is accurately reproduced for both high and low input
states. Instead the clock is phase shifted to a position where the
output accurately represents the input during a low state. During a
high state the output shows sharp spikes down to the low state. Thus
for transmitter input as shown in Figure~\ref{fig:xor} the result {\em
A} is obtained. To remedy this, an inverted copy of the phase shifted
clock is XOR combined with a second copy of the data to produce the
result {\em B}. The results {\em A} and {\em B} are used to clock a
dual data rate flip-flop at the output stage to reproduce the data,
{\em Q}. The behaviour of the flip-flop is given by the simple rules
in Table 1.

\section{Performance}
\begin{figure}[t]
\begin{center}
\includegraphics*[angle=-90, width=1.0\textwidth]{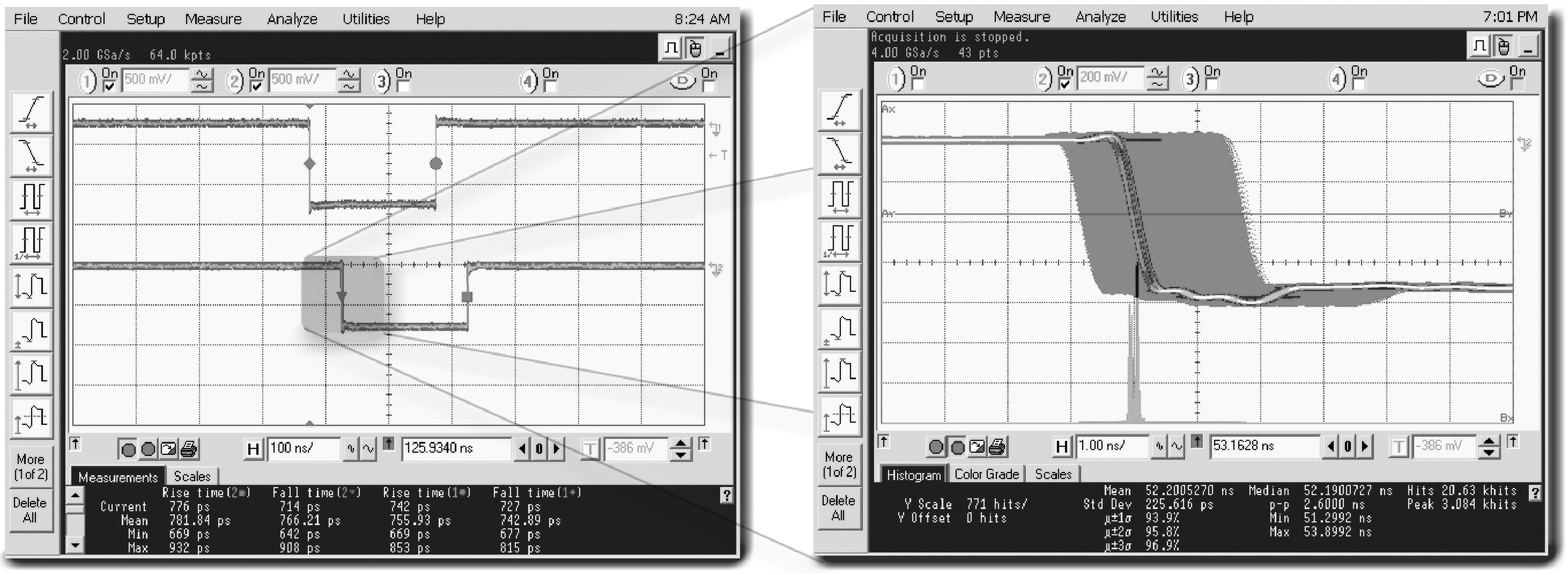}
\end{center}
\caption{Measurement of the arrival time of the falling edge of the DAT
output relative to a reference pulse (upper trace) for a single channel.}
\label{fig:screenshot}
\end{figure}

To characterise the performance of the DAT modules the arrival time of
the falling edge of a periodic, 1 MHz, 200 ns wide input to the
transmitter is measured at the receiver output, relative to the
falling edge of a reference pulse as shown in
Figure~\ref{fig:screenshot}. The arrival time varies about some mean
for a given channel. The maximum difference between the mean values of
the 11 channels is the skew. The variance in arrival times over many
measurements for a given channel is the jitter. The skew and average
jitter over all channels are taken as the two key indicators of
performance.

\begin{figure}[t]
\begin{center}
\includegraphics*[width=0.80\textwidth]{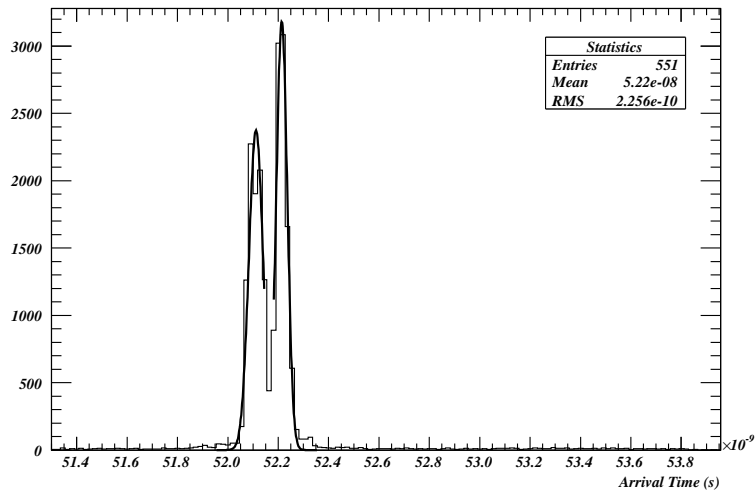}
\end{center}
\caption{Distribution of arrival times for a single channel at the DAT
output over a \mbox{2 m} fibre cable. Each peak is fitted with a
Gaussian to determine the random jitter of the channel. The
peak-to-peak value is used to evaluate the deterministic jitter of the
channel. }
\vspace{0.05in}
\label{fig:histogram}
\end{figure}

The term {\em jitter} simply refers to the uncertainty of a data edge
in time. There are several methods of measuring the jitter. The chosen
method here is to use an oscilloscope to record the arrival time of
the edge many times and add these values to a histogram. The jitter on
each channel is obtained from the distribution of around \mbox{20 k}
measurements over a \mbox{2 m} long, MPO terminated, 12 channel fibre
ribbon cable, as shown in Figure~\ref{fig:histogram}.

As the distribution of arrival times at the DAT output is not purely
Gaussian one cannot simply equate the jitter to the $RMS$ of the
distribution \cite{random,random2}. The total jitter is instead a
combination of two principle components: random jitter, $R_J$, and
deterministic jitter, $D_J$.

Random jitter is uncorrelated and unbounded. Given enough data
samples random jitter will have an unlimited peak-to-peak value. For
this reason $R_J$ is measured in terms of the standard deviation,
$\sigma_{R_J}$. The probability distribution function of random jitter is
always Gaussian. $R_J$ is often caused by thermal noise in decision
circuits and oscillator phase noise. The number of recorded data
samples is depends upon the probability for a logic transition
fluctuating across the sampling point, known as the bit error rate,
$BER$. The random jitter for a given $BER$ is related to the quantity
$Q$, a multiple of the standard deviation of the Gaussian, by
Eq.~\ref{eq:RJ}.

\begin{equation}
R_J = 2Q\sigma_{R_J}
\label{eq:RJ}
\end{equation}

Deterministic jitter is bounded and always measured in terms of a
peak-to-peak value. The distribution of deterministic jitter can be
very unpredictable. The total jitter, $T_J$ for a given BER is given
by the summation of $R_J$ and $D_J$ as shown in Eq.~\ref{eq:TJ}.

\begin{equation}
T_J = D_J + R_J = D_J + 2Q\sigma_{R_J}
\label{eq:TJ}
\end{equation}

The deterministic and random jitter may be separated from the recorded
distribution via the dual-Dirac model and used to predict a total
jitter for the industry standard BER of $10^{-12}$ (which corresponds
to $Q=7$) \cite{deterministic,deterministic2,deterministic3}. In the
dual-Dirac method the recorded distribution is modelled by two delta
functions displaced in time and convolved with a Gaussian. The
standard deviation of this Gaussian, $\sigma_{R_J}$, is estimated by
fitting the outer edges of the measured jitter distribution. The
displacement of the delta functions is given by the separation of the
outer peaks in the measured jitter distribution, and then taken to be
$D_J$.

The validity of this method was tested in two trials. First an
oscilloscope was set to average the waveform over a number of
acquisitions; the resulting jitter distribution averages out the
deterministic jitter, leaving only random jitter. The standard
deviation of this distribution proved consistent with $\sigma_{R_J}$
derived from the dual-Dirac method. Secondly the dual-Dirac model was
used to estimate $T_J$ for a BER of $10^{-7}$ and compared to the
peak-to-peak value of the measured jitter distribution for the same
BER. Again the two were found to be consistent. It should be noted
that it is not feasible to measure BERs of $<10^{-7}$ directly due to
the acquisition time required. The results of these trials indicate
that a sample of 20 k measurements is enough to adequately determine
both $D_J$ and $\sigma_{R_J}$, and therefore extrapolations to higher
BERs are valid.

A total jitter of 2.50 ns (or $\pm$1.25 ns) peak-to-peak for a BER of
$10^{-12}$ is obtained for the channel in the example distribution
shown in Figure~\ref{fig:screenshot}. The measurements for all
channels over a 60 m cable are shown in Table~\ref{table:tar} and an
average total jitter of 2.20 ns (or $\pm$1.10 ns) is obtained. This
BER extrapolation to 10$^{12}$ pulses is not intended as a replacement
for measurements made on a dedicated BER machine but simply provides a
quick estimation of the worst case jitter. The method provides an
estimate of the expected performance of a given channel and may be
used to rate channels according to performance for use with critical
signals within VERITAS. A more practical estimate of the jitter, in the
absence of a Gaussian distribution, that will effect the trigger
chain on an event-by-event basis is simply the time over which 95\%
of pulses arrive. On average over all channels, 95\% of pulses arrive
within 0.570 ns (or $\pm$0.265 ns) of the average for that
channel. This is crucially less than 1 ns, facilitating the accurate
transmission of VERITAS array trigger pulses. Consequently trigger
signals arrive at the FADC modules accurately enough to allow the
determination of the position of the data in the buffers to within one
FADC sample (2 ns), as described in Sections~\ref{Introduction} and
~\ref{Specification}. The skew between output channels from the
average arrival times is \mbox{1.6$\pm{0.4}$ ns} (at the 95\%
confidence level), small enough to integrate into the VERITAS trigger
chain without affecting the pre-programmed delays within the array
trigger system which compensate for the relative movement of the
telescopes.

Since the system introduces no deadtime the minimum pulse width that
can accurately be transmitted is only limited by the signal switching
speed and the system jitter. The minimum transmittable pulse width is
around 5 ns. This is also the time gap required between
consecutive pulses, and therefore the maximum transmittable data rate
is 200 MHz.

\section{Critical Evaluation}
The performance of the DAT modules exceeds the requirements of the
current VERITAS trigger chain. However, the FPGA code and methodology
have proven complicated due to the accuracy required to align the
clock with encoded data at the receiver. If the alignment is not
correct at the sub-nanosecond level a given channel will produce
spurious noise at either the clock frequency \mbox{(25 MHz)} or double
the clock frequency (50 MHz). These pulses have a width corresponding
directly to the misalignment of the clock and encoded data signal and
a minimum width of around \mbox{2 ns}, the signal switching time at
the FPGA output stage. Spurious pulses do not occur at all if a channel is
aligned.

The FPGA timing performance is adequate; all signals leaving the
transmitter are aligned to within 300 ps, but the duty cycle
distortion incurred is disappointing. The PAROLI modules also
introduce a 100 ps skew at the transmitter and again at the receiver.

In practice a fibre-ribbon cable must be used to connect the DAT
modules to ensure that the 12 fibres are of identical
length. Successful alignment over individual fibres is possible and
only requires that the fibres are of equal length to within 10 cm. However,
we found that when burying cable in 150 m of filled trenches on site
this proved challenging. Thus we turned to the fibre-ribbon cable as
the only viable outdoor option.

Laboratory tests showed no correlation between variations in either
circuit board or fibre temperature and the transit time and jitter of
pulses over approximately 10$^\circ$C. On site the modules are cooled
by fan trays to maintain an approximately constant temperature and
therefore the relatively narrow range of temperatures examined in the
laboratory is sufficient. The temperature of the fibre-optic cable
running between telescopes may experience more significant variations
in temperature than measured in the laboratory. The use of 12-channel
fibre-ribbon cable ensures that the transit times down both the clock
and data channels are effected equally by any extreme temperature
variations.

The load stability and the effect of instabilities in the power supply
have not been measured. It has been observed that switching modules
between VME crates with varying power supplies can affect the timing,
even knocking some channels out of alignment. This must be examined
further.

\section{Conclusion}
Fast digital trigger and event number distribution is achieved within
VERITAS using FPGA based DAT modules. The modules incorporate PAROLI
fibre optic interconnects to protect against lightning induced power
surges associated with coaxial cabling and to minimise
channel-to-channel skew and jitter. Combinatorial encoding and
decoding of the data results in a dead-time free system, and helps to
minimise the dead-time of the VERITAS array as a whole. The first
stereo data acquired by VERITAS (using two DAT pairs) has been
presented elsewhere \cite{Holder06b,Huges06,Frank06}. The four
telescope array requires 8 pairs of modules, all of which are
operational as of March 2007. A further 4 DAT pairs are to
be supplied as spares by October 2007. The DAT performance is limited by
the allowable channel-to-channel skew of the encoded data and clock
pulses arriving at the receiver. This currently requires a 12 channel
fibre-ribbon cable. Future design changes could correct for this
problem by introducing an FPGA with programmable delay on a individual
input pins.

This versatile technology could be adopted for other applications such
as triggers for particle physics experiments or clock distribution
within neutrino telescopes.

{\bf Acknowledgements}\\ The authors acknowledge the support of the
VERITAS collaboration and thank all those who helped with the
integration of the DAT modules into the detector. This work was made
possible with the financial support of the White Rose Studentship
programme and PPARC.

\clearpage
\bigskip
\begin{table}[h]
\begin{center}
\caption{Given inputs {\em A} and {\em B} the flip-flop, as configured
for the DAT receiver, will produce output {\em Q}. It is only
sensitive to rising edges ($\uparrow$). The symbol X indicates
the state of the input is irrelevant.}
\smallskip
\begin{tabular}{ccc} 
\hline\hline
A & B & Q \\ [0.5ex] 
\hline 
$\uparrow$   & X            & 1 \\ 
$\downarrow$ & X            & Q \\ [0.1ex]
X            & $\uparrow$   & 0 \\ [0.1ex]
X            & $\downarrow$ & Q \\ [1ex] 
\hline 
\end{tabular}
\end{center}
\label{table:flipflop}
\end{table}

\bigskip
\begin{table}[h] 
\begin{center}  
\caption{Arrival time statistics for all 11 data channels of the first
DAT pair over 60 m of fibre. Where $T_{Arrival}$ is the average
arrival time and $PP_{Arrival}$ and $RMS_{Arrival}$ are the
peak-to-peak and RMS values of the recorded
distribution. $\sigma_{R_J}$, $R_J$, $D_J$ and $T_J$ are the
determined standard deviation, random jitter, deterministic jitter,
and total jitter using the dual-Dirac model.}
\smallskip
\begin{tabular}{cccccccccc} 
\hline\hline 
Channel &$T_{Arrival}$ &$PP_{Arrival}$ &$RMS_{Arrival}$ &HITS    &$\sigma_{R_J}$   &$R_J$  &$D_J$     &$T_J$     &$95\%$\\[0.05ex]
        &$(ns)$        &$(ns)$         &$(ns)$ &$(kHits)$ &$(ns)$  &$(ns)$ &$(ns)$ &$(ns)$ &$(ns)$\\[0.5ex]
\hline
0  &346.00 &2.47 &0.22	&20.49	&0.03	&0.47	&2.22	&2.69	&1.02\\
1  &344.84 &1.55 &0.10	&20.28	&0.05	&0.73	&1.15	&1.88	&0.25\\
2  &344.45 &1.80 &0.16	&20.52	&0.03	&0.50	&1.53	&2.03	&0.40\\
3  &344.98 &1.73 &0.15	&20.17	&0.05	&0.67	&1.36	&2.03	&0.40\\
4  &345.71 &1.73 &0.13	&24.59	&0.03	&0.41	&1.50	&1.91	&0.40\\
5  &345.44 &1.58 &0.12	&20.86	&0.04	&0.56	&1.28	&1.84	&0.35\\
6  &345.48 &1.78 &0.16	&20.02	&0.04	&0.55	&1.48	&2.03	&0.47\\
7  &345.30 &2.02 &0.15	&20.74	&0.03	&0.50	&1.75	&2.24	&0.51\\
8  &345.10 &1.96 &0.17	&20.12	&0.02	&0.35	&1.78	&2.12	&0.69\\
9  &345.60 &3.13 &0.30	&20.97	&0.10	&1.44	&2.34	&3.78	&1.44\\
10 &345.14 &1.82 &0.12	&20.41	&0.03	&0.44	&1.58	&2.02	&0.36\\ [1ex]
\hline 
\vspace{0.05in}
\end{tabular}
\end{center}
\label{table:tar}
\end{table}
\bigskip

\clearpage

\end{document}